# Finding matched rms envelopes in rf linacs: A Hamiltonian approach

Robert D. Ryne
*MS H817, Accelerator Operations and Technology*
*Los Alamos National Laboratory*
*Los Alamos, NM 87545*

We present a new method for obtaining matched solutions of the rms envelope equations. In this approach, the envelope equations are first expressed in Hamiltonian form. The Hamiltonian defines a nonlinear mapping, $\mathcal{M}$, and for periodic transport systems the fixed points of the one-period map are the matched envelopes. Expanding the Hamiltonian around a fiducial trajectory, one obtains a linear map, $M$, that describes trajectories (rms envelopes) near the fiducial trajectory. Using $\mathcal{M}$ and $M$ we construct a contraction mapping that can be used to obtain the matched envelopes. The algorithm is quadratically convergent. Using the zero-current matched parameters as starting values, the contraction mapping typically converges in a few to several iterations. Since our approach uses numerical integration to obtain all the mappings, it includes the effects of nonidealized, $z$-dependent transverse and longitudinal focusing fields. We present numerical examples including finding a matched beam in a quadrupole channel with rf bunchers.

## INTRODUCTION

In 1959 Kapchinskij and Vladimirskij published the first envelope equations governing two dimensional beams with space charge [1]. Although they assumed an unusual phase space distribution (a $\delta$-function whose argument was a linear function of the Courant-Snyder invariants) their result was important because it described, in terms of a set of ordinary differential equations, the self-consistent transport of finite emittance beams in strong focusing systems. More than 10 years later Sacherer and Lapostolle separately showed that, for beams with elliptical symmetry, one could derive rms envelope equations that were satisfied by beams in general and not just KV beams [2,3]. They showed that for linear external fields (but no such restriction on the beam self-fields) one could obtain a set of equations that involved only second moments, but to achieve this they had to allow the beam emittance to appear in the rms equations as an unknown function of time. Sacherer also showed that one could derive envelope equations for three dimensional beams (*i.e.* a six dimensional phase space) assuming ellipsoidal symmetry; unlike the two dimensional case, the equations depend on the precise form of the distribution through a parameter called $\lambda_3$, but that dependence turns out to be very weak. Through the years activity has continued in the area of envelope equations. Motivated by a desire to analyze the transport of intense electron beams in gas, Lee and Cooper derived rms equations for cylindrically symmetric beams including scattering; additionally, their formulation included nonzero canonical angular momentum and acceleration [4]. In order to model beam transport systems and rf linacs, Crandall developed the now widely used codes TRACE and TRACE3D [5]. More recently, Struckmeier derived envelope equations starting from the Fokker-Planck equation [6]. He used the formalism to estimate emittance growth due to intra-beam scattering in storage rings.

In this paper we will emphasize the application of envelope equations to quadrupole channels and rf linacs. Envelope equations have turned out to be extremely useful in the early stages of the design of these systems. Our approach allows a very accurate treatment of the beamline elements, since it takes into account $z$-dependent longitudinal and transverse fields of rf gaps and focusing magnets. The fields can be in the form of measured field data or analytic functions that approximate field data.

Often one is interested not only in numerically integrating the rms equations but also in finding initial conditions that result in periodic solutions when the beam transport system is itself periodic. In this paper we will present an efficient method for finding these matched solutions. Our method rests on the fact that the rms equations can be represented as a Hamiltonian system. While this system's nonlinear behavior can be found through numerical integration of the rms equations, its linear behavior can be obtained using standard techniques in accelerator physics for computing linear maps. Together, the linear and nonlinear maps can be used to construct a quadratically convergent iterative procedure for finding matched rms envelopes. Typically, we have found the procedure to converge in a few to several iterations, even under conditions of extreme space charge depression.

# I. TWO DIMENSIONAL SYSTEMS

## A. Overview

Consider a particle beam propagating in a quadrupole channel. Suppose that the beam is long compared with its transverse dimensions, and that we can neglect any longitudinal variation when calculating the beam self-fields. We will neglect image charge effects, and we will suppose that the beam is launched along the axis of a perfectly aligned transport system. We will use the longitudinal coordinate, $z$, as the independent variable. The canonical coordinates and momenta for the transverse phase space are denoted $(x, p_x, y, p_y)$. Let the vector potential associated with the quadrupoles be given by

$$A_x = A_y = 0, \tag{1}$$

$$A_z = \frac{1}{2} g(z)(y^2 - x^2), \tag{2}$$

where $g(z)$ denotes the magnetic quadrupole gradient. Let $\Psi^{\rm self}$ denote the scalar potential associated with the self-fields and, neglecting transverse currents, suppose that the associated vector potential is given by

$$A_x = A_y = 0, \tag{3}$$

$$A_z = \frac{\beta_o}{c} \Psi^{\rm self}, \tag{4}$$

where $\beta_o c$ is the velocity on the design trajectory. Rather than working with the variables $(x, p_x, y, p_y)$ it is convenient to define dimensionless variables $(\bar{x}, \bar{p}_x, \bar{y}, \bar{p}_y)$ according to

$$\bar{x} = x/l, \quad \bar{p}_x = p_x/p_o, \tag{5}$$

$$\bar{y} = y/l, \quad \bar{p}_y = p_y/p_o, \tag{6}$$

where $p_o$ denotes the momentum on the design trajectory (*i.e.* $p_o = \gamma_o \beta_o mc$) and where $l$ is a scale length [7]. The Hamiltonian (in MKSA units) governing these variables is given approximately by

$$H(\bar{x}, \bar{p}_x, \bar{y}, \bar{p}_y, z) = \frac{1}{2l}(\bar{p}_x^2 + \bar{p}_y^2)$$

$$+ \frac{lk(z)}{2}(\bar{x}^2 - \bar{y}^2) + \frac{K/2}{l} \hat{\Psi}(l\bar{x}, l\bar{y}, z), \tag{7}$$

where

$$k(z) = (q/p_o) g(z), \tag{8}$$

and where $K$ is the generalized perveance,

$$K = \frac{qI}{2\pi\epsilon_o p_o v_o^2 \gamma_o^2}. \tag{9}$$

Also, $\hat{\Psi}$ is related to $\Psi^{\rm self}$ according to

$$\Psi^{\rm self} = \frac{\lambda}{4\pi\epsilon_o} \hat{\Psi}, \tag{10}$$

where $\lambda$ is the charge per unit length measured in the lab frame, $\lambda = I/v_o$. Note that we have expanded the relativistic Hamiltonian to second order in the phase space variables, with the exception of the scalar potential, as is the standard procedure for deriving rms envelope equations. For the remainder of the discussion of two dimensional systems we will set $l = 1$ m.

Following the usual procedure one can obtain equations for the rms envelopes, $X$ and $Y$ [2,3]. The envelope equations are given by

$$\frac{d^2X}{dz^2} + kX - \frac{K/2}{X+Y} - \frac{\mathcal{E}_x^2}{X^3} = 0,$$

$$\frac{d^2Y}{dz^2} - kY - \frac{K/2}{X+Y} - \frac{\mathcal{E}_y^2}{Y^3} = 0, \tag{11}$$

where $\mathcal{E}_x$ and $\mathcal{E}_y$ denote unnormalized rms emittances. Since these are rms equations the factor in the space charge term is $K/2$, whereas it would be $2K$ for the KV equations. The envelope equations are derivable from a Hamiltonian, $H^{\text{env}}$, where

$$H^{\text{env}}(X,P_x,Y,P_y)=\frac{1}{2}(P_x^2+\frac{\mathcal{E}_x^2}{X^2})+\frac{1}{2}(P_y^2+\frac{\mathcal{E}_y^2}{Y^2})+\frac{k}{2}(X^2-Y^2)-(K/2)\log(X+Y). \tag{12}$$

Lastly, one can use the envelope Hamiltonian to define depressed phase advances, $\mu_x$ and $\mu_y$, of a particle in an equivalent KV beam. Referring to the envelope Hamiltonian, we regard the phase advances as coordinates and the emittances as a canonically conjugate momenta [8]. (Since the phase advance appears nowhere in the Hamiltonian, the emittance is constant, as expected). Taking this view, we obtain

$$\mu_x=\mathcal{E}_x\int\frac{dz}{X^2},$$
$$\mu_y=\mathcal{E}_y\int\frac{dz}{Y^2}. \tag{13}$$

For a KV distribution, these formulas apply to all the particles in the beam (since the forces are linear). For other distributions they apply to the equivalent KV beam (*i.e.* a KV beam with the same rms values). The phase advance equations can be integrated along with the envelope equations.

### B. Constructing the contraction map

Consider a periodic transport system with period $L$. Let $\zeta=(X,P_x,Y,P_y)$. As shown above the envelope equations are derivable from a Hamiltonian; hence they define a symplectic nonlinear mapping $\mathcal{M}$ that maps initial state vectors into final state vectors:

$$\zeta^{\text{fin}}=\mathcal{M}\zeta^{\text{in}}. \tag{14}$$

If we consider transport through one period of the transport system, then a matched envelope is simply a fixed point of $\mathcal{M}$:

$$\mathcal{M}\zeta=\zeta. \tag{15}$$

Techniques for finding fixed points of symplectic maps are widely used in accelerator physics [9]. For example, they are used to find the off-momentum closed orbits of particles in circular machines. In our case, however, we will use the techniques to find the fixed points of an envelope map, not a particle map. The approach is based on the fact that the machinery exists to compute the action of the nonlinear map $\mathcal{M}$ as well as its linear part, $M$. This makes it easy to construct a contraction map based on a Newton search procedure to find the fixed point. As a result, the method is quadratically convergent. As an illustration, consider the problem of finding solutions of the equation $g(x)=x$, or equivalently, of finding roots of the function $f(x)=g(x)-x$. The Newton search algorithm defines a contraction mapping $C$ that, for sufficiently close starting values, converges to a root [10]. If $x^n$ is the value of $x$ on the nth iteration, then applying the contraction map $C$ to $x^n$ produces a value at the next iteration:

$$x^{n+1}=Cx^n=x^n-\frac{f(x^n)}{f'(x^n)}=x^n-\frac{x^n-g(x^n)}{1-g'(x^n)}. \tag{16}$$

It is easily shown that if $x^n$ is within $\epsilon$ of a root then $Cx^n$ deviates by an amount proportional to $\epsilon^2$.

For a multidimensional system, the contraction map is given by [9]

$$\zeta^{n+1}=C\zeta^n=\zeta^n-(I-M)^{-1}(\zeta^n-\mathcal{M}\zeta^n), \tag{17}$$

where $I$ is the identity matrix and $M$ is the matrix associated with linear part of $\mathcal{M}$. To obtain $M$, we first need to linearize $H^{env}$ about a "given" fiducial trajectory, $\zeta_g$. Let $\hat{\zeta}=\zeta-\zeta_g$. The quadratic part of the Hamiltonian governing these deviation variables, which we will denote $H_2$, is given by

$$H_2=\frac{1}{2}(\hat{P}_x^2+\hat{P}_y^2)+\frac{\hat{X}^2}{2}[k+\frac{3\mathcal{E}_x^2}{X_g^4}+\frac{(K/2)}{(X_g+Y_g)^2}]+\frac{\hat{Y}^2}{2}[-k+\frac{3\mathcal{E}_y^2}{Y_g^4}+\frac{(K/2)}{(X_g+Y_g)^2}]+\frac{(K/2)\hat{X}\hat{Y}}{(X_g+Y_g)^2}. \tag{18}$$

The equation of motion for $M$ is well known [11]:

$$\frac{dM}{dz} = JSM, \tag{19}$$

where the symmetric matrix $S$ is related to $H_2$ by

$$H_2(\hat{\zeta}, t) = \frac{1}{2} \sum_{a,b=1}^{4} S_{ab}\hat{\zeta}_a\hat{\zeta}_b, \tag{20}$$

and where the matrix $J$ is the fundamental symplectic two-form (*i.e.* the nonzero elements of $J$ are given by $J_{12} = J_{34} = 1$, $J_{21} = J_{43} = -1$). Comparing Eq. (18) and Eq. (20) (with $\hat{\zeta} = (\hat{X}, \hat{P}_x, \hat{Y}, \hat{P}_y)$) one can immediately identify the matrix elements of $S$. That is, $S_{11}$ is the coefficient of $\frac{1}{2}\hat{X}^2$, $S_{22}$ is the coefficient of $\frac{1}{2}\hat{P}_x^2$, $S_{13} = S_{31}$ is the coefficient of $\hat{X}\hat{Y}$, *etc.*

Summarizing, Eq. (17) defines a contraction map for finding matched rms envelopes. In order to evaluate the right hand side of the equation, one must use numerical integration to compute the following: (1)$\mathcal{M}\zeta$, which is just the numerical solution of Hamilton's equations with the Hamiltonian of Eq. (12); and (2)the matrix $M$, which is obtained by numerically integrating Eq. (19). These quantities are computed at every iteration until the difference between $C\zeta$ and $\zeta$ is sufficiently small. In the calculations below we consider the map to have converged when $|\zeta - C\zeta|/|\zeta| < 10^{-8}$. We use the zero current matched values as starting values. For systems where these cannot be found analytically, we begin by integrating the envelope equations with zero current and zero emittance; this is equivalent to computing the beta functions, from which we obtain the zero current matched envelopes.

### C. Example: matching in a quadrupole channel

As a numerical example, consider a beam in a magnetic quadrupole channel of the "FODO" type. The quadrupoles in this example are idealized as having a constant gradient over their length, though since numerical integration is used anyway, it would be little or no extra work to use analytic models or numerical values for the quadrupole gradients. The period length $L = 24$ cm and each quadrupole is 6 cm long. The channel is designed to transport a beam of 10 MeV protons having rms emittances $\mathcal{E}_x = \mathcal{E}_y = 1 \times 10^{-6}$ m-rad. The zero current phase shift per cell is 60 degrees, which requires $g = 78$ T/m. At zero current, the matched rms beam sizes are given by $X = 0.6310$ mm and $Y = 0.3796$ mm. With these as starting values the contraction map converges in 5 iterations when the beam current is $I = 8.4$ amp, as is shown in the following computer output listing:

```
Enter current, x-emittance, y-emittance:
  8.4   1.e-6   1.e-6

Zero current matched rms values:
X=6.3103d-04, P_x=-2.0322d-19
Y=3.7961d-04, P_y=-3.3781d-19
Zero current phase shifts per cell:
sigma0_x=59.971, sigma0_y=59.971

Starting contraction mapping...
iteration 1: delta=  3.326906d-00
iteration 2: delta=  8.062686d-01
iteration 3: delta=  3.867283d-03
iteration 4: delta=  1.059674d-04
iteration 5: delta=  7.206193d-10
search converged

Matched rms values:
X=1.9492d-03, P_x=1.45795d-10
Y=1.2217d-03, P_y=2.42433d-10
Phase shifts per cell:
sigma_x=6.011, sigma_y=6.011
```

Note that the convergence is very rapid even though the depressed phase advance is only 6 degrees per cell.

## II. THREE DIMENSIONAL SYSTEMS

### A. Overview

Consider a particle beam propagating in a beamline consisting of quadrupole magnets and cylindrically symmetric rf cavities. These elements make it possible to approximately model structures such as drift tube linacs and coupled cavity linacs. The potentials associated with the quadrupole magnets and the beam self-fields are the same is in the two dimensional case (but we will use the notation $g_m$ instead of $g$ to denote the magnetic quadrupole gradient). Additionally, the vector potential associated with an rf cavity is of the form

$$\begin{aligned} A_x &= \frac{e'(z)}{2\omega_\alpha} x \sin(\omega_\alpha t + \theta) \\ A_y &= \frac{e'(z)}{2\omega_\alpha} y \sin(\omega_\alpha t + \theta) \\ A_z &= -\frac{1}{\omega_\alpha}\{e(z) - \frac{r^2}{4}[e''(z) + \frac{\omega_\alpha^2}{c^2} e(z)]\} \sin(\omega_\alpha t + \theta), \end{aligned} \tag{21}$$

where the electric field at $r = 0$ is given by

$$E_z(r = 0) = e(z)\cos(\omega_\alpha t + \theta), \tag{22}$$

and where a prime denotes $d/dz$.

As before, we use the longitudinal coordinate $z$ as the independent variable. Now there are six canonical coordinates and momenta, denoted $(x, p_x, y, p_y, t, p_t)$, where $t$ denotes a particle's arrival time at the location $z$ and where its canonically conjugate momentum $p_t$ is the negative of its total energy. As above, we will define variables that are dimensionless deviations from the design trajectory. First, let $(t_o, p_{to})$ denote the design trajectory (along with $x = p_x = y = p_y = 0$). Using the above potentials, it follows that the equations of motion for the design trajectory are given by

$$t_o' = \frac{-p_{to}/c}{\sqrt{p_{to}^2 - m^2c^4}}, \tag{23}$$

$$p_{to}' = -qe(z)\cos(\omega_\alpha t + \theta). \tag{24}$$

Note that, in the two dimensional case, the momentum on the design orbit was a constant, and it was natural to scale the transverse momenta by this quantity. But in the three dimensional case, the rf fields can accelerate the beam. Thus, we will scale the transverse momenta by a parameter $\delta$ which is unrelated to the design momentum. (In fact, later it will be convenient to set $\delta = mc$). The transverse coordinates are scaled by a parameter $l$. Lastly, the time $t$ will be scaled by a quantity $\omega$, so that the times are really phases. Often one would choose $\omega = \omega_\alpha$, but due to the frequency changes typical of proton or ion linacs, there are instances where one would choose $n\omega = \omega_\alpha$, where $n$ is an integer. In summary, the dimensionless deviation variables are given by

$$\bar{x} = x/l, \quad \bar{p}_x = p_x/\delta \tag{25}$$

$$\bar{y} = y/l, \quad \bar{p}_y = p_y/\delta \tag{26}$$

$$\bar{t} = \omega(t - t_o), \quad \bar{p}_t = (p_t - p_{to})/(\omega l \delta) \tag{27}$$

The single particle Hamiltonian, $H(\bar{x}, \bar{p}_x, \bar{y}, \bar{p}_y, \bar{t}, \bar{p}_t, z)$, paraxial in the external fields, is given by

$$\begin{aligned} H = &\frac{\delta}{2lp_o}(\bar{p}_x^2 + \bar{p}_y^2) + \frac{qlg_m}{2\delta}(\bar{x}^2 - \bar{y}^2) + \frac{l}{2\delta}\left[\frac{1}{p_o}\left(\frac{q}{2\omega_\alpha}e'\sin\phi_s\right)^2 - \frac{q}{2\omega_\alpha}\left(e'' + \frac{\omega_\alpha^2}{c^2}e\right)\sin\phi_s\right](\bar{x}^2 + \bar{y}^2) \\ &-\frac{qe'\sin\phi_s}{2p_o\omega_\alpha}(\bar{x}\bar{p}_x + \bar{y}\bar{p}_y) + \frac{m^2\omega^2 l\delta}{2p_o^3}\bar{p}_t^2 - \frac{\omega_\alpha qe\sin\phi_s}{2\omega^2 l\delta}\bar{t}^2 + \frac{p_o}{l\delta}(\bar{K}/2)\hat{\Psi}(l\bar{x}, l\bar{y}, \bar{t}/\omega + t_o, z), \end{aligned} \tag{28}$$

where the synchronous phase, $\phi_s$, is given by

$$\phi_s = \omega_\alpha t_o(z) + \theta. \tag{29}$$

In the above Hamiltonian, $\hat{\Psi}$ is obtained from Eq. (10) with $\lambda = \bar{I}/v_o$, and $\bar{K}$ is the perveance obtained from Eq. (9) using the average current $\bar{I}$. Note that if the scaling parameter $\omega$ is chosen to be the frequency of the bunches, then the charge per bunch $Q$ is related to $\bar{I}$ according to $Q = (2\pi/\omega)\bar{I}$. In what follows, it will be convenient to define a quantity $u_o$, where

$$u_o = \frac{\gamma_o\beta_o}{\omega l/c}. \tag{30}$$

Following Sacherer [2], one can obtain equations for the rms envelopes, $X, Y$ and $T$:

$$\begin{aligned}
&X'' + \frac{p_o'}{p_o}X' + \frac{qg_m}{p_o}X - \frac{qg_{rf}}{p_o}X - \frac{\bar{K}u_o\pi\lambda_3}{l^2}XG_{311}(X,Y,u_oT) - (\frac{\delta}{lp_o})^2\frac{\mathcal{E}_{n,x}^2}{X^3} = 0,\\
&Y'' + \frac{p_o'}{p_o}Y' - \frac{qg_m}{p_o}Y - \frac{qg_{rf}}{p_o}Y - \frac{\bar{K}u_o\pi\lambda_3}{l^2}YG_{131}(X,Y,u_oT) - (\frac{\delta}{lp_o})^2\frac{\mathcal{E}_{n,y}^2}{Y^3} = 0,\\
&T'' + 3\frac{p_o'}{p_o}T' - \frac{q\omega_\alpha/c^2}{\gamma_o^2\beta_o^2p_o}e\sin\phi_s T - \frac{\bar{K}u_o\pi\lambda_3}{l^2}TG_{113}(X,Y,u_oT) - (\frac{\delta}{lp_ou_o^2})^2\frac{\mathcal{E}_{n,t}^2}{T^3} = 0,
\end{aligned} \tag{31}$$

where

$$g_{rf} = \frac{1}{2}[\frac{\omega_\alpha}{c^2}e\sin\phi_s - \frac{(\omega/\omega_\alpha)^2}{v_o}e'\cos\phi_s]. \tag{32}$$

In the above equations, $\mathcal{E}_{n,x}$, $\mathcal{E}_{n,y}$ and $\mathcal{E}_{n,t}$ denote normalized rms emittances. The quantity $\lambda_3$ is a geometrical factor that depends on the details of the charge distribution within the bunch, but as Sacherer pointed out it is not very sensitive to the details and has a value approximately equal to $1/(5\sqrt{5})$ for a wide variety of distributions. Lastly, the quantity $G$ is a space charge term defined by

$$G_{mnp}(x,y,z) = \frac{3}{2}\int_0^\infty \frac{ds}{(x^2+s)^{m/2}(y^2+s)^{n/2}(z^2+s)^{p/2}}. \tag{33}$$

Note that these rms equations are *not* expected to accurately model the bunching process. The reason for this is twofold: (1)By our paraxial expansion of the single particle Hamiltonian, we assume that the external rf fields vary linearly across a bunch; (2)The space charge terms are based on the fields of an *isolated* bunch of charge, not a train of bunches.

Lastly, the rms equations are derivable from the following envelope Hamiltonian:

$$\begin{aligned}
H^{\text{env}}(X,P_x,Y,P_y,T,P_t) &= \frac{\delta}{2lp_o}(P_x^2 + \frac{\mathcal{E}_{n,x}^2}{X^2}) + \frac{\delta}{2lp_o}(P_y^2 + \frac{\mathcal{E}_{n,y}^2}{Y^2}) + \frac{\delta}{2lp_ou_o^2}(P_t^2 + \frac{\mathcal{E}_{n,t}^2}{T^2})\\
&+\frac{qlg_m}{2p_o}(X^2-Y^2) - \frac{qlg_{rf}}{2\delta}(X^2+Y^2) - \frac{q\omega_\alpha e\sin\phi_s}{2\omega^2 l\delta}T^2 + \frac{\bar{K}u_o^2\pi\lambda_3}{l}G_{111}(X,Y,u_oT).
\end{aligned} \tag{34}$$

### B. Constructing the contraction map

As in the two dimensional case, we need to compute the matrix $M$ that describes the linear behavior of the system governed by $H^{\text{env}}$. This in turn requires that we know the matrix $S$, which appears in Eqs. (19) and (20). Linearizing $H^{\text{env}}$ around a given fiducial trajectory $\zeta_g$ we obtain the following nonzero matrix elements of $S$:

$$S_{22} = S_{44} = \frac{\delta}{lp_o}, \quad S_{66} = \frac{\delta}{lp_ou_o^2}, \tag{35}$$

$$S_{11} = \frac{ql}{\delta}(g_m - g_{rf}) + \frac{\delta}{lp_o}\frac{3\mathcal{E}_{n,x}^2}{X_g^4} + \frac{\bar{K}u_o^2\pi\lambda_3}{l}(3X_g^2G_{511} - G_{311}), \tag{36}$$

$$S_{33} = \frac{ql}{\delta}(-g_m - g_{rf}) + \frac{\delta}{lp_o}\frac{3\mathcal{E}_{n,y}^2}{Y_g^4} + \frac{\bar{K}u_o^2\pi\lambda_3}{l}(3Y_g^2G_{151} - G_{131}), \tag{37}$$

$$S_{55} = -\frac{q\omega_\alpha e \sin\phi_s}{\omega^2 l\delta} + \frac{\delta}{lp_o}\frac{3\mathcal{E}_{n,t}^2}{u_o^2 T_g^4} + \frac{\bar{K}u_o^4\pi\lambda_3}{l}(3u_o^2T_g^2G_{115} - G_{113}), \tag{38}$$

$$S_{13} = S_{31} = \frac{\bar{K}u_o^2\pi\lambda_3}{l}X_gY_gG_{331}, \tag{39}$$

$$S_{15} = S_{51} = \frac{\bar{K}u_o^4\pi\lambda_3}{l}X_gT_gG_{313}, \tag{40}$$

$$S_{35} = S_{53} = \frac{\bar{K}u_o^4\pi\lambda_3}{l}Y_gT_gG_{133}. \tag{41}$$

Note that we have used the notation $G_{mnp}$ to denote $G_{mnp}(X, Y, u_oT)$ in the above equations.

The following section contains a numerical example of finding matched beams in a three dimensional system. But before continuing we need to specify our choice of scaling parameters:

$$n\omega = \omega_\alpha, \tag{42}$$

$$\delta = mc, \tag{43}$$

$$l = c/\omega. \tag{44}$$

In the above equations, $\omega$ is simply the frequency of the bunches, normally a harmonic of the rf frequency. If every rf bucket is filled, $n = 1$; if every other bucket is filled, $n = 2$; and so on. By choosing $\omega l/c = 1$, it means that $X$ and $Y$ need to be multiplied by the inverse wavenumber, $k^{-1} = c/\omega$, to convert them to dimensional quantities. Lastly, note that with this choice of parameters the phase advances are given by

$$\begin{aligned}
\mu_x &= \frac{\mathcal{E}_{n,x}}{l}\int\frac{dz}{\gamma_o\beta_o X^2}, \\
\mu_y &= \frac{\mathcal{E}_{n,y}}{l}\int\frac{dz}{\gamma_o\beta_o Y^2}, \\
\mu_t &= \frac{\mathcal{E}_{n,t}}{l}\int\frac{dz}{\gamma_o^3\beta_o^3 T^2}.
\end{aligned} \tag{45}$$

### C. Example: matching in a quadrupole channel with bunchers

We will consider the same FODO channel as described previously, but in addition rf gaps will be inserted between each quadrupole. This is shown schematically in Fig. 1. Though we could use numerical field data or analytic functions that approximate field data, for the sake of illustration we have assumed that $e(z)$ is a sum of two identical, longitudinally separated Gaussians:

$$e(z) = E_{max}(e^{(z-z_1)^2/2\sigma^2} + e^{(z-z_2)^2/2\sigma^2}), \tag{46}$$

with $z_1 = 6$ cm, $z_2 = 18$ cm, $\sigma = 5$ mm and $E_{max} = 20$ MV/m. Each cavity is assumed to have the same frequency and phase (see Eqs. (21)): $f_\alpha = \omega_\alpha/2\pi = 361.75$ MHz, $\theta = 90.06$ degrees. With these choices of frequency, phase, and cavity separation, a synchronous particle crosses the gaps with a phase of roughly $-90$ degrees, the net result being no acceleration or deceleration. Fig. 2 shows the synchronous particle's energy as a function of $z$.

The transverse emittances are chosen to be the same as in the two dimensional example. Since we now need dimensionless normalized values, we have to take the numbers from the two dimensional example and multiply them by a factor $(\gamma_o\beta_o/l)$. This factor equals 1.11, and so the resulting dimensionless normalized rms emittances are given by $\mathcal{E}_{n,x} = \mathcal{E}_{n,y} = 1.11 \times 10^{-6}$. For simplicity this is also the value we choose for $\mathcal{E}_{n,t}$. As before, the channel is designed to transport a beam of 10 MeV protons. At zero current, the matched rms beam sizes are $X = 4.968 \times 10^{-3}$, $Y = 3.009 \times 10^{-3}$ and $T = 3.416 \times 10^{-2}$ rad. (In physical units, the transverse values are $X = 0.6553$ mm and $Y = 0.3969$ mm, since $l = 0.1319$ m). At zero current, the rf defocusing effect of the gaps depresses the transverse phase advances from 60 degrees to around 55 degrees per period; the temporal (*i.e.* longitudinal) phase advance is slightly more than 30 degrees per period. At a beam current of 150 ma, the contraction map converges in 6 iterations:

```
 freq,curr,x-emit,y-emit,t-emit (normalized):
361.75e6  0.150  1.11e-6  1.11e-6  1.11e-6

Zero current matched rms values:
X=4.9679d-03, P_x=-2.0090d-10
Y=3.0094d-03, P_y=6.1963d-12
T=3.4158d-02, P_t=-2.7831d-12
Zero current phase shifts per cell:
sig0_x=55.377,sig0_y=55.376,sig0_t=31.434

Starting contraction mapping...
iteration 1: delta=  9.460663d-01
iteration 2: delta=  4.247934d-01
iteration 3: delta=  3.944055d-02
iteration 4: delta=  5.912194d-04
iteration 5: delta=  1.848957d-06
iteration 6: delta=  4.322956d-09
search converged

Matched rms values:
X=8.5011d-03, P_x=1.1776d-09
Y=5.3076d-03, P_y=2.4236d-09
T=9.9245d-02, P_t=1.9463d-12
Phase shifts per cell:
sig_x=18.301,sig_y=18.300,sig_t=3.724
```

Fig. 3 shows the matched rms envelopes in the transverse directions, and Fig. 4 shows the matched temporal envelope.

## SUMMARY

The purpose of this paper has been to present a method for finding matched rms envelopes in beamlines consisting of quadrupole magnets and cylindrically symmetric rf gaps. We did this by casting the envelope equations in Hamiltonian form and constructing a contraction map based on a Newton search procedure. Our numerical examples show that the map converges very rapidly even under conditions of significant space charge depression.

---

[1] I.M. Kapchinskij and V.V. Vladimirskij, Proceedings of the International Conference on High Energy Accelerators and Instrumentation, CERN, Geneva, p. 274 (1959).

[2] F. Sacherer, IEEE Trans. Nuc. Sci. NS-18, No. 3, 1105 (1971).

[3] P. Lapostolle, IEEE Trans. Nuc. Sci. NS-18, No. 3, 1101 (1971). In fact Lapostolle's treatment of two dimensional beams required only two-fold symmetry, with elliptical symmetry as a special case. The general form of his equations contain additional distribution-dependent terms that are usually small.

[4] E.P. Lee and R.K. Cooper, *Particle Accelerators*, **7**, 83 (1976)

[5] K.R. Crandall and D.P. Rusthoi, *Documentation for TRACE: An Interactive Beam Transport Code*, Los Alamos National Laboratory Report LA-10235-MS; K.R. Crandall and D.P. Rusthoi, TRACE 3-D Documentation, Second Edition, Los Alamos National Laboratory Report LA-UR-90-4146 (December 1990).

[6] J. Struckmeier, *Particle Accelerators*, (1994).

[7] One should not confuse the scale length $l$ with other quantities such at the cell length $L$ in periodic systems or the quadrupole length in "$gl$" products; $l$ is simply a length by which the coordinates are scaled to make them dimensionless. Often it is chosen to be 1 m, in which case the coordinates, expressed as dimensionless quantities, are really expressed in meters. In some situations (such as the three dimensional systems described in the text) other choices of $l$ are useful.

[8] R.D. Ryne, Ph.D. Thesis, University of Maryland (1987).

[9] A.J. Dragt, *Lectures on Nonlinear Orbit Dynamics*, in American Institute of Physics Conference Proceedings No. 87, edited by R.A. Carrigan, F.R. Huson and M. Month (AIP, New York, 1982).

[10] W. Press, S. Teukolsky, W. Vetterling and B. Flannery, Numerical Recipes in FORTRAN, Cambridge University Press (1992).

[11] A.J. Dragt and E. Forest, *J. Math. Phys.* **24**, 2734 (1983).

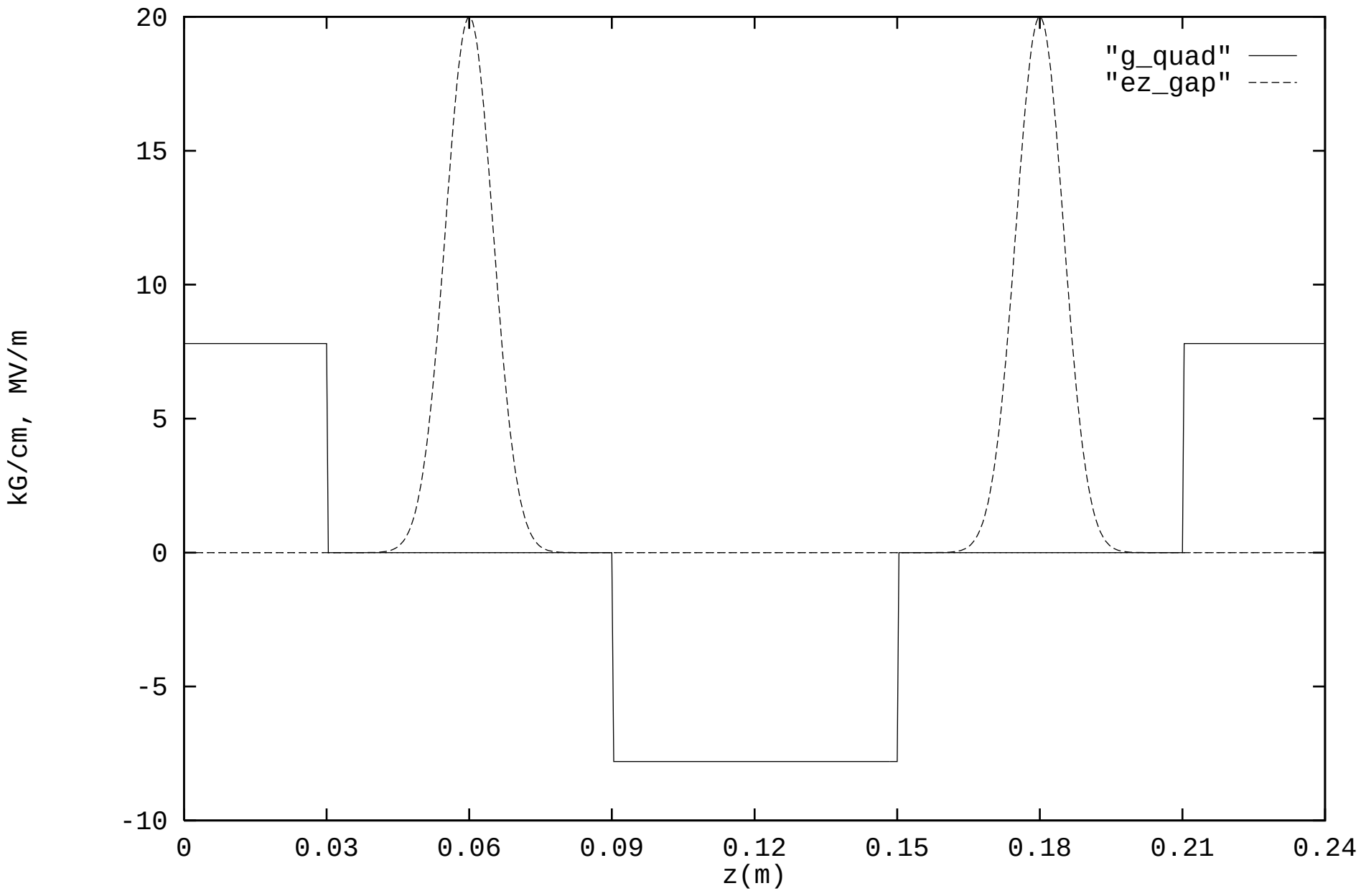

FIG. 1. One cell of a quadrupole channel with rf bunchers, showing quadrupole gradient (solid line) and longitudinal electric field (dashed line).

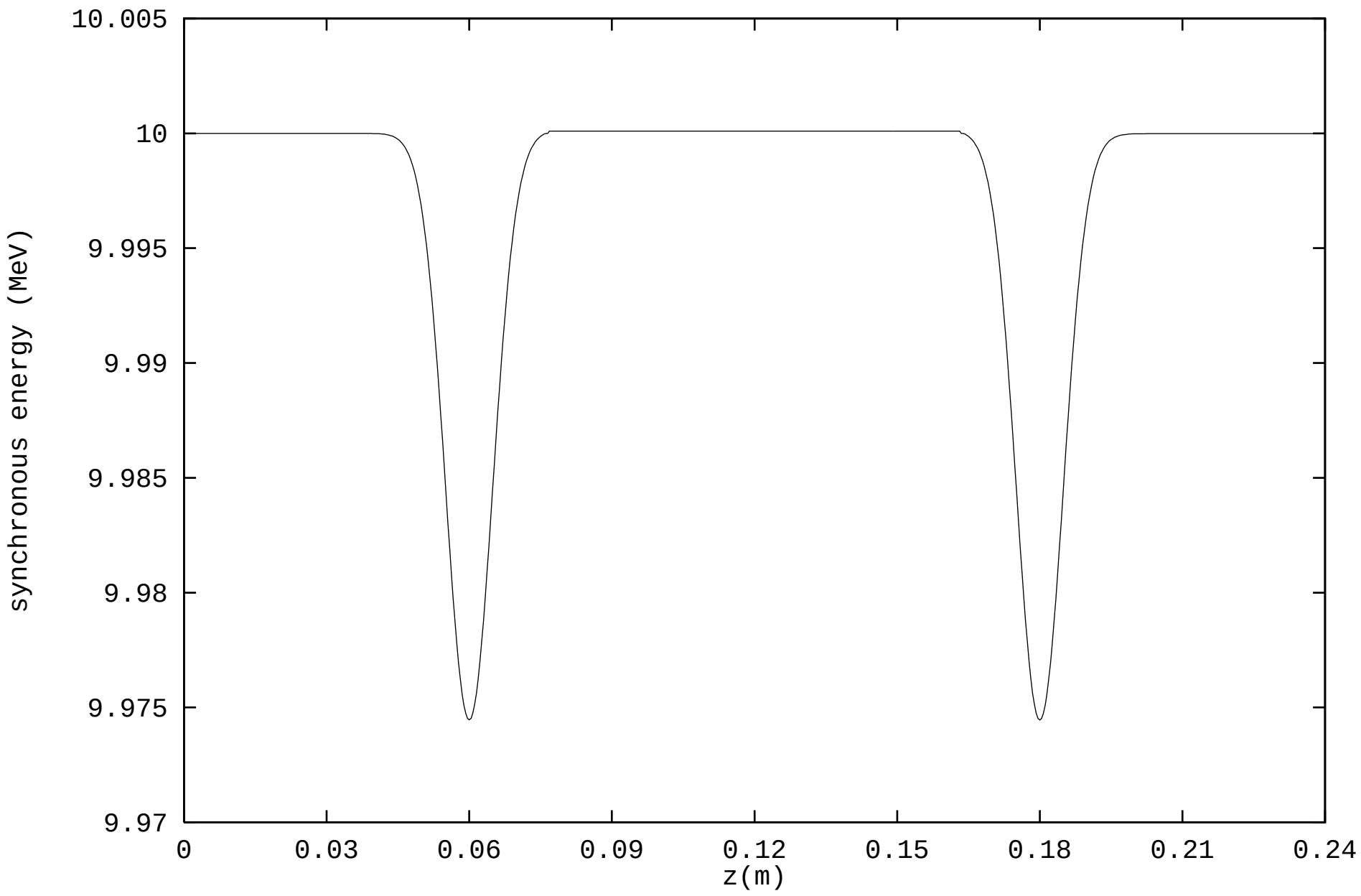

FIG. 2. Energy versus $z$ for a synchronous particle propagating in one cell of the quadrupole channel with bunchers.

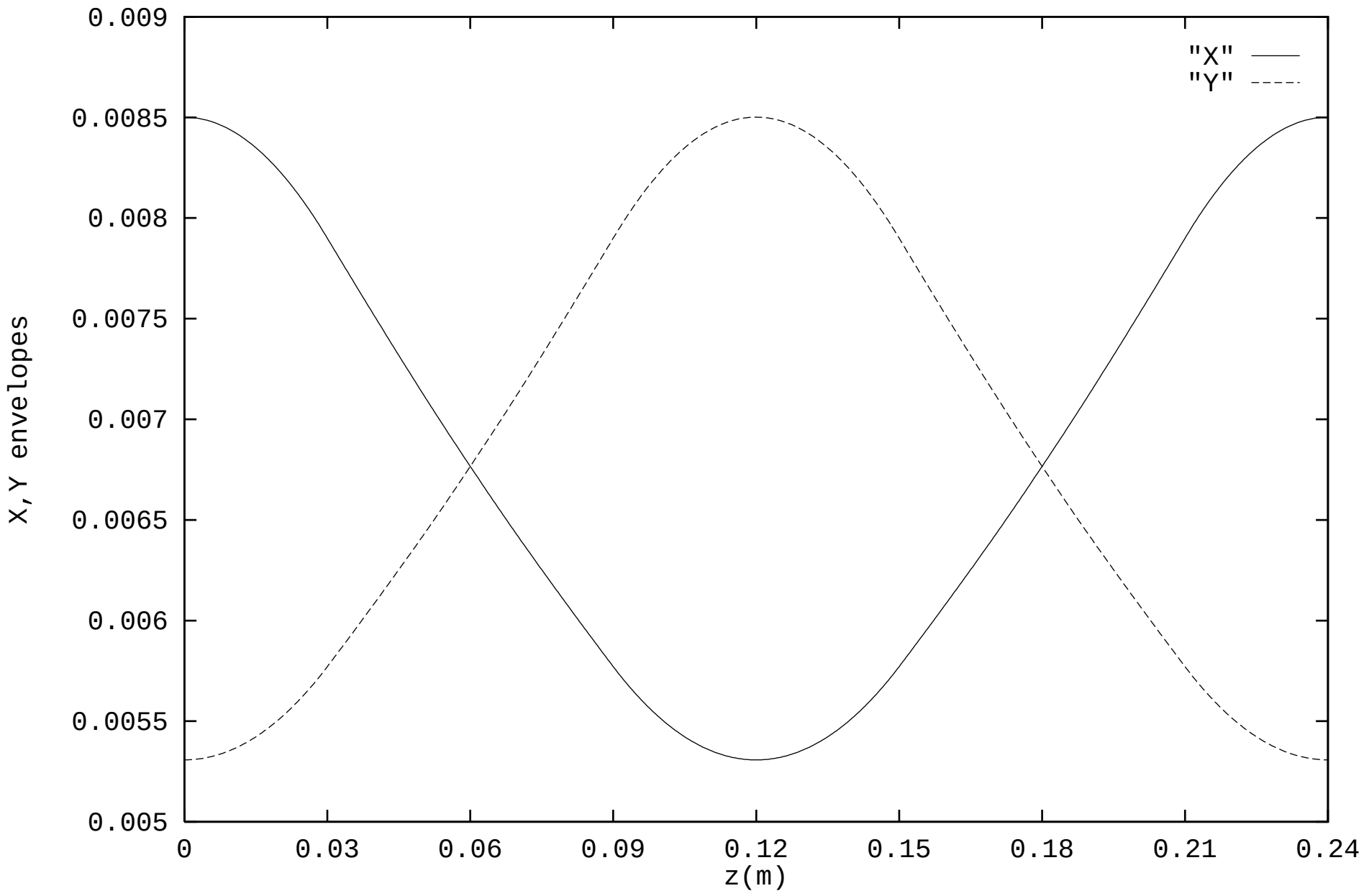

FIG. 3. Matched transverse envelopes for the quadrupole channel with bunchers. Since $X$ and $Y$ are dimensionless, they must be multiplied by the scale length $l = 0.1319$ m to convert them to meters.

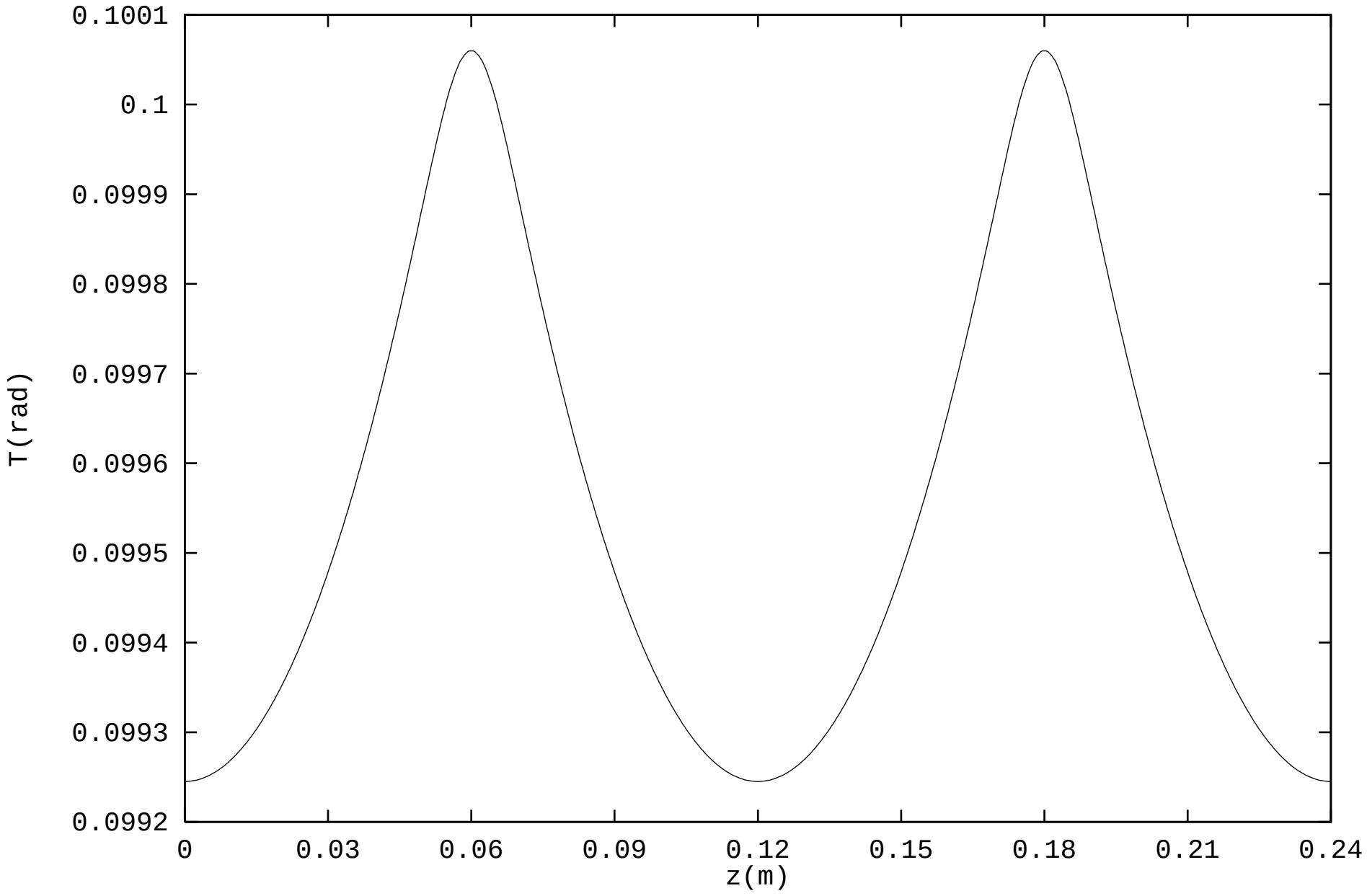

FIG. 4. Matched temporal (*i.e.* longitudinal) envelope $T$ for the quadrupole channel with bunchers.